\documentclass[a4paper, 10 pt, conference]{ieeeconf}
\IEEEoverridecommandlockouts
\usepackage{amsfonts,amsmath,amssymb} 

\usepackage{psfrag,color}
\usepackage{enumerate,cite,latexsym,graphicx}
\newtheorem{theorem}{Theorem}
\newtheorem{lemma}{Lemma}

\newtheorem{definition}{Definition}

\newtheorem{assumption}{Assumption}
\def\tr{\mathop{\rm Tr}\nolimits} 

\title{\LARGE \bf Robust Stability of Uncertain Quantum Systems
}

\author{Ian R.~Petersen, Valery Ugrinovskii and Matthew R~James  %
\thanks{This work was supported by the
Australian Research Council (ARC) and Air Force Office of Scientific
Research (AFOSR). This material is based on research sponsored by the
Air Force Research Laboratory, under agreement number
FA2386-09-1-4089.  The U.S. Government is authorized to reproduce and
distribute reprints for Governmental purposes notwithstanding any
copyright notation thereon.
The views and conclusions contained herein are those of the authors
and should not be interpreted as necessarily representing the official
policies or endorsements, either expressed or implied, of the Air
Force Research Laboratory or the U.S. Government. }%
\thanks{Ian R. Petersen and Valery Ugrinovskii are with the School of  Engineering and Information Technology, 
        University of New South Wales at the Australian Defence Force Academy, Canberra ACT 2600, Australia.
         {\tt\small \{i.r.petersen,v.ugrinovskii\}@gmail.com} } 
\thanks{Matthew R. James is with the Research School of  Engineering,
College of Engineering and Computer Science, The Australian National University, Canberra, ACT 0200,
Australia. Email: Matthew.James@anu.edu.au.}
}%

\begin{document}

\maketitle
\thispagestyle{empty}
\pagestyle{empty}

\begin{abstract}
This paper considers the problem of robust stability for a class of
uncertain quantum systems subject to unknown perturbations in the
system Hamiltonian. Some general stability results are given for
different classes of perturbations to the system Hamiltonian. Then, the
special case of a nominal linear quantum system is considered with
either quadratic or non-quadratic perturbations to the system
Hamiltonian. In this case, robust stability conditions are given in terms of strict bounded real conditions. 
\end{abstract}

\section{Introduction} \label{sec:intro}
An important concept in modern control theory is the notion of
robust or absolute stability for uncertain nonlinear systems in the form of a
Lur'e system with an uncertain nonlinear block which satisfies a sector bound
condition; e.g., see \cite{KHA02}. This enables a frequency domain
condition for robust stability to be given. This
characterization of robust stability enables robust feedback
controller synthesis to be carried out using $H^\infty$ control
theory; e.g., see \cite{ZDG96}. The aim of this paper is to  extend
classical results on  robust  stability to the case of
quantum systems. This is motivated by a desire to apply quantum
$H^\infty$ control such as presented in \cite{JNP1,MaP4} to nonlinear
and uncertain quantum systems.

In recent years, a number of papers have considered the feedback
control of systems whose dynamics are governed by the laws of quantum
mechanics rather than classical mechanics; e.g., see
\cite{YK03A,YK03B,YAM06,JNP1,NJP1,GGY08,MaP3,MaP4,YNJP1,GJ09,GJN10,WM10,PET10Ba}. In
particular, the papers \cite{GJ09,JG10} consider a framework of
quantum systems defined in terms of a triple $(S,L,H)$ where $S$ is a
scattering matrix, $L$ is a vector of coupling operators and $H$ is a
Hamiltonian operator. The paper \cite{JG10} then introduces notions of
dissipativity and stability for this class of quantum systems. In this
paper, we build on the results of \cite{JG10} to obtain robust 
stability results for uncertain quantum systems in which the quantum system Hamiltonian is
decomposed as $H =H_1+H_2$ where $H_1$ is a known nominal Hamiltonian
and $H_2$ is a perturbation Hamiltonian, which is contained in a
specified set of Hamiltonians $\mathcal{W}$. The set of perturbation
Hamiltonians $\mathcal{W}$ corresponds to the set of exosystems
considered in \cite{JG10}.

For this general class of uncertain quantum systems, a number of stability
results are obtained. The paper then considers the case in which the
nominal Hamiltonian $H_1$ is a quadratic function of annihilation and
creation operators and the coupling operator vector is a linear
function of annihilation and creation operators. This case corresponds
to a nominal  linear quantum system; e.g., see
\cite{JNP1,NJP1,MaP3,MaP4,PET10Ba}. In this special case, robust stability
results are obtained in terms of a frequency domain condition. 

The remainder of the paper proceeds as follows. In Section
\ref{sec:systems}, we define the general class of uncertain quantum
systems under consideration. In Section \ref{sec:quad_pert}, we
consider a special class of quadratic perturbation Hamiltonians and
obtain a robust stability result for this case. In Section
\ref{sec:nonquadratic}, we consider a general class of  non-quadratic perturbation Hamiltonians. In Section
\ref{sec:linear}, we specialize to the case of a linear nominal
quantum systems and obtain a number of robust stability results for this case
in which stability conditions are given in terms of a strict bounded
real condition. In Section \ref{sec:example}, we
present an illustrative example and in Section \ref{sec:conclusions},
we present some conclusions. 

\section{Quantum Systems} \label{sec:systems}
We consider  open quantum systems defined by  parameters $(S,L,H)$ where $H = H_1+H_2$; e.g., see \cite{GJ09,JG10}.  The corresponding generator for this quantum system is given by 
\begin{equation}
\label{generator}
\mathcal{G}(X) = -i[X,H] + \mathcal{L}(X)
\end{equation}
where $ \mathcal{L}(X) = \frac{1}{2}L^\dagger[X,L]+\frac{1}{2}[L^\dagger,X]L$. Here, $[X,H] = XH-HX$ denotes the commutator between two operators and the notation $^\dagger$ denotes the adjoint transpose of a vector of operators. Also, $H_1$ is a self-adjoint operator on the underlying Hilbert space referred to as the nominal Hamiltonian and $H_2$ is a self-adjoint operator on the underlying Hilbert space referred to as the perturbation Hamiltonian.  The triple $(S,L,H)$, along with the corresponding generators define the Heisenberg evolution $X(t)$ of an operator $X$ according to a quantum stochastic differential equation; e.g., see \cite{JG10}.

\begin{figure}[htbp]
\begin{center}
\includegraphics[width=6cm]{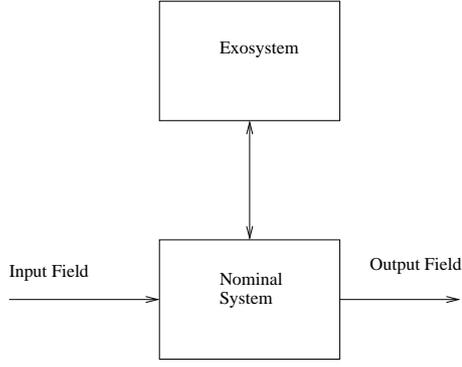}
\end{center}
\caption{Block diagram representation of an open quantum system interacting with an exosystem.}
\label{F1}
\end{figure}

The problem under consideration involves establishing robust stability
properties for an uncertain open quantum system for the case in which the perturbation Hamiltonian is contained in a given set $\mathcal{W}$. Using the notation of \cite{JG10}, the set $\mathcal{W}$ defines a set of exosystems. This situation is illustrated in the block diagram shown in Figure \ref{F1}. 
The main robust stability results presented in this paper will build on the following result from \cite{JG10}. 
\begin{lemma}[See Lemma 3.4 of \cite{JG10}.]
\label{L0}
Consider an open quantum system defined by $(S,L,H)$ and suppose there exists a non-negative self-adjoint operator $V$ on the underlying Hilbert space such that
\begin{equation}
\label{lyap_ineq}
\mathcal{G}(V) + cV \leq \lambda
\end{equation}
where $c > 0$ and $\lambda$ are real numbers. Then for any plant state, we have
\[
\left<V(t)\right> \leq e^{-ct}\left<V\right> + \frac{\lambda}{c},~~\forall t \geq 0.
\]
Here $V(t)$ denotes the Heisenberg evolution of the operator $V$ and $\left<\cdot\right>$ denotes quantum expectation; e.g., see \cite{JG10}.
\end{lemma}

\subsection{Commutator Decomposition}
Given set of non-negative self-adjoint operators $\mathcal{P}$ and  real parameters $\gamma > 0$, $\delta \geq 0$, we now define a particular set of perturbation Hamiltonians $\mathcal{W}_1$. This set $\mathcal{W}_1$ is defined in terms of the commutator decomposition
\begin{equation}
\label{comm_condition}
[V,H_2] = [V,z^\dagger]w-w^\dagger[z,V]
\end{equation}
for  $V \in \mathcal{P}$ where $w$ and $z$ are given vectors of
operators. Here, the notation $[z,V]$ for a vector of operators $z$
and a scalar operator $V$ denotes the corresponding vector of
commutators. Also, this set will be defined in terms of the sector bound condition:
\begin{equation}
\label{sector1}
w^\dagger w \leq \frac{1}{\gamma^2}z^\dagger z + \delta.
\end{equation}
Indeed, we define
\begin{equation}
\label{W1}
\mathcal{W}_1 = \left\{\begin{array}{l}H_2: \exists w,~z \mbox{ such that (\ref{sector1}) is satisfied} \\
\mbox{ and (\ref{comm_condition}) is satisfied } \forall V \in \mathcal{P}\end{array}\right\}.
\end{equation}
Using this definition, we obtain the following theorem. 
\begin{theorem}
\label{T1}
Consider a set of non-negative self-adjoint operators $\mathcal{P}$ and an open quantum system $(S,L,H)$ where $H=H_1+H_2$ and $H_2 \in \mathcal{W}_1$ defined in (\ref{W1}). If there exists a $V \in \mathcal{P}$ and real constants $c > 0$, $\tilde \lambda \geq 0$ such that 
\begin{equation}
\label{dissip}
-i[V,H_1]+ \mathcal{L}(V)+ [V,z^\dagger][z,V] + \frac{1}{\gamma^2}z^\dagger z + cV \leq \tilde \lambda,
\end{equation}
then
\[
\left<V(t)\right> \leq e^{-ct}\left<V\right> + \frac{\tilde \lambda+\delta}{c},~~\forall t \geq 0.
\]
\end{theorem}

\noindent
{\em Proof:}
Let $V \in \mathcal{P}$ be given and consider $\mathcal{G}(V)$ defined in (\ref{generator}). Then
\begin{eqnarray}
\label{ineq1}
\mathcal{G}(V) &=&  -i[V,H_1]+ \mathcal{L}(V) -i[V,z^\dagger]w+iw^\dagger[z,V] \nonumber \\
\end{eqnarray}
using (\ref{comm_condition}).
Now since $V$ is self-adjoint $[V,z^\dagger]^\dagger= [z,V]$. Therefore, 
\begin{eqnarray*}
0 &\leq& \left([V,z^\dagger]- iw^\dagger\right)\left([V,z^\dagger]- iw^\dagger\right)^\dagger\nonumber \\
&=&  [V,z^\dagger][z,V]+i[V,z^\dagger]w-iw^\dagger[z,V]+w^\dagger w.
\end{eqnarray*}
Substituting this into (\ref{ineq1}), it follows that
\begin{eqnarray}
\label{ineq2}
\mathcal{G}(V) 
& \leq & -i[V,H_1]+ \mathcal{L}(V)+ [V,z^\dagger][z,V] + \frac{1}{\gamma^2}z^\dagger z + \delta \nonumber \\
\end{eqnarray}
using (\ref{sector1}). 
 Hence, (\ref{dissip}) implies (\ref{lyap_ineq}) holds with $\lambda = \tilde \lambda + \delta$.
Therefore, the result follows from Lemma \ref{L0}.
\hfill $\Box$

\subsection{Alternative Commutator Decomposition}
We now consider an alternative set of perturbation Hamiltonians
$\mathcal{W}_2$. For a given set of non-negative self-adjoint operators
$\mathcal{P}$ and real parameters $\gamma > 0$, $\delta_1\geq 0$, $\delta_2\geq 0$,  this set $\mathcal{W}_2$ is defined in terms of the commutator decomposition
\begin{eqnarray}
\label{alt_comm_condition}
[V,H_2] &=& [V,z]w_1^*-w_1[z^*,V]\nonumber \\
&&+\frac{1}{2}\left[z,[V,z]\right]w_2^*-\frac{1}{2}w_2\left[z,[V,z]\right]^*
\end{eqnarray}
for  $V \in \mathcal{P}$ where $w_1$, $w_2$ and $z$ are given scalar
operators. Here, the notation $^*$ denotes the adjoint of an
operator. Also, the set $\mathcal{W}_2$ will be defined in terms of the
sector bound condition
\begin{equation}
\label{sector2a}
w_1 w_1^* \leq \frac{1}{\gamma^2}z z^* + \delta_1
\end{equation}
and the condition
\begin{equation}
\label{sector2b}
w_2 w_2^* \leq \delta_2.
\end{equation}
Then, we define
\begin{equation}
\label{W2}
\mathcal{W}_2 = \left\{\begin{array}{l}H_2: \exists w_1,~w_2,~z \mbox{
      such that 
(\ref{sector2a}) and (\ref{sector2b}) } \\
\mbox{ are satisfied and (\ref{alt_comm_condition}) is satisfied } \forall V \in \mathcal{P}\end{array}\right\}.
\end{equation}
Using this definition, we obtain the following theorem. 
\begin{theorem}
\label{T2}
Consider a set of non-negative self-adjoint operators $\mathcal{P}$
and an open quantum system $(S,L,H)$ where $H=H_1+H_2$ and $H_2 \in
\mathcal{W}_2$ defined in (\ref{W2}). If there exists a $V \in
\mathcal{P}$ and real constants  $c > 0$,
$\tilde \lambda \geq 0$ such that 
$\mu = \left[z,[V,z]\right]$ is a constant and\begin{eqnarray}
\label{dissip1a}
&&-i[V,H_1]+ \mathcal{L}(V)
+ [V,z][z^*,V]
+ \frac{1}{ \gamma^2}zz^*
+ cV \leq \tilde \lambda.
\nonumber \\
\end{eqnarray}
Then 
\[
\left<V(t)\right> \leq e^{-ct}\left<V\right> + \frac{\tilde \lambda+ \delta_1+\mu\mu^*/4+\delta_2}{c},~~\forall t \geq 0.
\]
\end{theorem}

\noindent
{\em Proof:}
Let $V \in \mathcal{P}$ be given and consider $\mathcal{G}(V)$ defined in (\ref{generator}). Then 
\begin{eqnarray}
\label{ineq1a}
\mathcal{G}(V) &=& -i[V,H_1]+ \mathcal{L}(V) -i[V,z]w_1^*+iw_1[z^*,V]\nonumber \\
&&-i\mu w_2^*+iw_2\mu^*
\end{eqnarray}
using (\ref{alt_comm_condition}).
Now $[V,z]^* = z^*V-Vz^*=[z^*,V]$ since $V$ is self-adjoint. Therefore, 
\begin{eqnarray*}
0 &\leq& \left([V,z]- i w_1\right)
\left([V,z]- iw_1\right)^*\nonumber \\
&=&  [V,z][z^*,V]+i[V,z]w_1^*\nonumber \\
&&-iw_1[z^*,V]+w_1 w_1^*
\end{eqnarray*}
and hence
\begin{eqnarray}
\label{ineq3a}
-i[V,z]w_1^*+iw_1[z^*,V] \leq [V,z][z^*,V]+w_1w_1^*.
\end{eqnarray}

Also, 
\begin{eqnarray*}
0 &\leq& \left(\frac{1}{2}\mu- i w_2\right)
\left(\frac{1}{2}\mu- i w_2\right)^*\nonumber \\
&=&  \frac{1}{4}\mu\mu^*-\frac{i}{2} w_2\mu^*+\frac{i}{2}\mu w_2^*+ w_2 w_2^*
\end{eqnarray*}
and hence
\begin{eqnarray}
\label{ineq3b}
\frac{i}{2}w_2\mu^*-\frac{i}{2}\mu w_2^* &\leq& \frac{1}{4}\mu\mu^*
+w_2 w_2^*.
\end{eqnarray}

Substituting (\ref{ineq3a}) and (\ref{ineq3b}) into (\ref{ineq1a}), it follows that
\begin{eqnarray}
\label{ineq2a}
\mathcal{G}(V) &\leq &  -i[V,H_1]+ \mathcal{L}(V)+ [V,z][z^*,V]
\nonumber \\&&
+\frac{1}{\gamma^2} z z^* 
+\delta_1
+\mu\mu^*/4+\delta_2
\end{eqnarray}
using (\ref{sector2a}) and (\ref{sector2b}). 
 Then it follows from (\ref{dissip1a}) that 
\[
\mathcal{G}(V) + cV \leq \tilde \lambda + \delta_1+\mu\mu^*/4+\delta_2. 
\]
Then the result of the theorem follows from Lemma \ref{L0}.
\hfill $\Box$

\section{Quadratic Perturbations of the Hamiltonian}
\label{sec:quad_pert}
In this section, we consider a set $\mathcal{W}_3$ of quadratic perturbation
Hamiltonians of the following form
\begin{equation}
\label{H2quad}
H_2 = \frac{1}{2}\left[\begin{array}{cc}\zeta^\dagger &
      \zeta^T\end{array}\right]\Delta
\left[\begin{array}{c}\zeta \\ \zeta^\#\end{array}\right]
\end{equation}
where $\Delta \in \mathbb{C}^{2m\times 2m}$ is a Hermitian matrix of the
form
\begin{equation}
\label{Delta_form}
\Delta= \left[\begin{array}{cc}\Delta_1 & \Delta_2\\
\Delta_2^\# &     \Delta_1^\#\end{array}\right]
\end{equation}
and $\Delta_1 = \Delta_1^\dagger$, $\Delta_2 = \Delta_2^T$. Also,
$\zeta = E_1a+E_2 a^\#$. Here $a$ is a vector of annihilation
operators on the underlying Hilbert space and $a^\#$ is the
corresponding vector of creation operators. Also, in the case of matrices, the notation $^\dagger$ refers to the complex conjugate transpose of a matrix. In the case vectors of
operators, the notation $^\#$ refers to the vector of adjoint
operators and in the case of complex matrices, this notation refers to
the complex conjugate matrix. 

The annihilation and creation operators are assumed to satisfy the
canonical commutation relations:
\begin{eqnarray}
\label{CCR2}
\left[\left[\begin{array}{l}
      a\\a^\#\end{array}\right],\left[\begin{array}{l}
      a\\a^\#\end{array}\right]^\dagger\right]
&=&\left[\begin{array}{l} a\\a^\#\end{array}\right]
\left[\begin{array}{l} a\\a^\#\end{array}\right]^\dagger
\nonumber \\
&&- \left(\left[\begin{array}{l} a\\a^\#\end{array}\right]^\#
\left[\begin{array}{l} a\\a^\#\end{array}\right]^T\right)^T\nonumber \\
&=& J
\end{eqnarray}
where $J = \left[\begin{array}{cc}I & 0\\
0 & -I\end{array}\right]$; e.g., see \cite{GGY08,GJN10,PET10Ba}.

The matrix $\Delta$
is subject to the norm bound
\begin{equation}
\label{Delta_bound}
\|\Delta\| \leq \frac{2}{\gamma}
\end{equation}
where $\|\cdot\|$ denotes the matrix induced norm (maximum singular
value). 
Then we define
\begin{equation}
\label{W3}
\mathcal{W}_3 = \left\{\begin{array}{l}H_2 \mbox{ of the form
      (\ref{H2quad}) such that 
} \\
\mbox{ conditions (\ref{Delta_form} and (\ref{Delta_bound}) are satisfied}\end{array}\right\}.
\end{equation}
Using this definition, we obtain the following lemma. 
\begin{lemma}
\label{LA}
For any set of self-adjoint operators $\mathcal{P}$,
\[
\mathcal{W}_3 \subset \mathcal{W}_1.
\]
\end{lemma}

\noindent
{\em Proof:}
Given any $H_2 \in \mathcal{W}_3$, let 
\[
w = \frac{1}{2}\left[\begin{array}{cc}\Delta_1 & \Delta_2\\
\Delta_2^\# &     \Delta_1^\#\end{array}\right]\left[\begin{array}{c}\zeta \\ \zeta^\#\end{array}\right]
= \frac{1}{2}\left[\begin{array}{c}\Delta_1\zeta+\Delta_2\zeta^\# \\\Delta_2^\# \zeta+ \Delta_1^\#\zeta^\#\end{array}\right]
\]
and 
\begin{equation}
\label{z0}
z = \left[\begin{array}{c}\zeta \\ \zeta^\#\end{array}\right]= 
\left[\begin{array}{cc}
E_1 & E_2\\
E_2^\# & E_1^\#
\end{array}\right]\left[\begin{array}{c}a \\ a^\#\end{array}\right] 
=E\left[\begin{array}{c}a \\ a^\#\end{array}\right].
\end{equation}
Hence, 
\[
H_2 = w^\dagger z = \frac{1}{2}\left[\begin{array}{cc}a^\dagger &
      a^T\end{array}\right]E^\dagger \Delta E \left[\begin{array}{c}a \\ a^\#\end{array}\right].
\]
Then, for any  $V \in \mathcal{P}$,
\begin{eqnarray*}
[V,z^\dagger]w 
&=& \frac{1}{2}\left(\begin{array}{c}V\zeta^\dagger \Delta_1\zeta+V\zeta^\dagger \Delta_2 \zeta^\#\\+V\zeta^T\Delta_2^\# \zeta+    V\zeta^T \Delta_1^\# \zeta^\# \end{array}\right) \nonumber \\
&&- \frac{1}{2}\left(\begin{array}{c}\zeta^\dagger V \Delta_1\zeta+ \zeta^\dagger V\Delta_2\zeta^\#\\+\zeta^TV\Delta_2^\# \zeta+    \zeta^T V\Delta_1^\# \zeta^\#
\end{array}\right).
\end{eqnarray*}
Also, 
\begin{eqnarray*}
w^\dagger[z,V]
&=&
\frac{1}{2}\left(\begin{array}{c}\zeta^\dagger \Delta_1\zeta V +\zeta^T\Delta_2^\# \zeta V\\+\zeta^\dagger \Delta_2\zeta^\#V+ \zeta^T\Delta_1^\#\zeta^\#V\end{array}\right) \nonumber \\
&-& \frac{1}{2}\left(\begin{array}{c}\zeta^\dagger V\Delta_1\zeta +\zeta^TV\Delta_2^\# \zeta\\+\zeta^\dagger V\Delta_2\zeta^\#+ \zeta^TV\Delta_1^\#\zeta^\#\end{array}\right).
\end{eqnarray*}
Hence, 
\begin{eqnarray*}
\lefteqn{[V,z^\dagger]w - w^\dagger[z,V]} \nonumber \\
 &=& \frac{1}{2}\left(V\zeta^\dagger \Delta_1\zeta
+V\zeta^\dagger \Delta_2 \zeta^\#+V\zeta^T\Delta_2^\#\zeta+    V\zeta^T \Delta_1^\# \zeta^\# \right) \nonumber \\
& -&\frac{1}{2}\left(\begin{array}{c}\zeta^\dagger \Delta_1\zeta V
+\zeta^T\Delta_2^\# \zeta V\\+\zeta^\dagger \Delta_2\zeta^\#V + \zeta^T\Delta_1^\#\zeta^\#V\end{array}\right) \nonumber \\
&=& VH_2-H_2V 
=[V,H_2]
\end{eqnarray*}
and thus (\ref{comm_condition}) is satisfied. Also,  condition (\ref{Delta_bound}) implies
\[
\frac{1}{4}\left[\begin{array}{cc}\zeta^\dagger &
      \zeta^T\end{array}\right]\Delta\Delta\left[\begin{array}{c}\zeta \\ \zeta^\#\end{array}\right] \leq 
\frac{1}{\gamma^2}\left[\begin{array}{cc}\zeta^\dagger &
      \zeta^T\end{array}\right]\left[\begin{array}{c}\zeta \\ \zeta^\#\end{array}\right]
\]
which implies (\ref{sector1}) for any $\delta \geq 0$. Hence, $H_2 \in
\mathcal{W}_1$. Since, $H_2 \in
\mathcal{W}_3$ was arbitrary, we must have $\mathcal{W}_3 \subset \mathcal{W}_1$.
\hfill $\Box$

\section{Non-quadratic Perturbation Hamiltonians}
\label{sec:nonquadratic}
In this section, we define a  set of non-quadratic perturbation
Hamiltonians denoted $\mathcal{W}_4$.   For a given set of non-negative self-adjoint operators
$\mathcal{P}$ and real parameters $\gamma > 0$,  $\delta_1\geq 0$, $\delta_2\geq 0$, 
the set $\mathcal{W}_4$ is defined in terms of the following  power series (which is assumed to converge in the sense of the induced operator norm on the underlying Hilbert space)
\begin{equation}
\label{H2nonquad}
H_2 = f(\zeta,\zeta^*) = \sum_{k=0}^\infty\sum_{\ell=0}^\infty S_{k\ell}\zeta^k(\zeta^*)^\ell = \sum_{k=0}^\infty \sum_{\ell=0}^\infty S_{k\ell} H_{k\ell}.
\end{equation}
Here $S_{k\ell}=S_{\ell k}^*$, $H_{k\ell} = \zeta^k(\zeta^*)^\ell$, and $\zeta$ is a scalar operator on the underlying Hilbert space. It follows from this definition that
\[
H_2^* = \sum_{k=0}^\infty\sum_{\ell=0}^\infty S_{k\ell}^*\zeta^\ell(\zeta^*)^k = 
\sum_{\ell=0}^\infty\sum_{k=0}^\infty S_{\ell k}\zeta^\ell(\zeta^*)^k = H_2
\]
and thus $H_2$ is a self-adjoint operator. Note that it follows from
the use of Wick ordering that the form (\ref{H2nonquad}) is the most
general form for a perturbation Hamiltonian defined in terms of a
single scalar operator $\zeta$. 

Also, we let 
\begin{equation}
\label{fdash}
f'(\zeta,\zeta^*) = \sum_{k=1}^\infty\sum_{\ell=0}^\infty k S_{k \ell} \zeta^{k-1}(\zeta^*)^\ell,
\end{equation}
\begin{equation}
\label{fddash}
f''(\zeta,\zeta^*) = \sum_{k=1}^\infty\sum_{\ell=0}^\infty k(k-1)S_{k\ell} \zeta^{k-2}(\zeta^*)^{\ell}
\end{equation}
and consider the 
sector bound condition
\begin{equation}
\label{sector4a}
f'(\zeta,\zeta^*)^*f'(\zeta,\zeta^*)  \leq \frac{1}{\gamma^2}\zeta \zeta^* + \delta_1
\end{equation}
and the condition
\begin{equation}
\label{sector4b}
f''(\zeta,\zeta^*)^*f''(\zeta,\zeta^*) \leq  \delta_2.
\end{equation}
Then we define the set $\mathcal{W}_4$  as follows:
\begin{equation}
\label{W5}
\mathcal{W}_4 = \left\{\begin{array}{l}H_2 \mbox{ of the form
      (\ref{H2nonquad}) such that 
} \\
\mbox{ conditions (\ref{sector4a}) and (\ref{sector4b}) are satisfied}\end{array}\right\}.
\end{equation}
Note that the condition (\ref{sector4b}) effectively amounts to a global Lipschitz condition on the quantum nonlinearity. 

In this section, the set of non-negative self-adjoint operators
$\mathcal{P}$ will be assumed to satisfy  the following  assumption:
\begin{assumption}
\label{A1}
Given any $V \in \mathcal{P}$, the quantity
\[
\mu = \left[\zeta,[V,\zeta]\right] = \zeta [V,\zeta]-[V,\zeta]\zeta
\]
is a constant.
\end{assumption}

\begin{lemma}
\label{LB}
Suppose the set of self-adjoint operators $\mathcal{P}$ satisfies
Assumption \ref{A1}. Then
\[
\mathcal{W}_4 \subset \mathcal{W}_2.
\]
\end{lemma}

\noindent
{\em Proof:}
First, we note that given any $V \in \mathcal{P}$ and $k \geq 1$,
\begin{eqnarray}
\label{Vzetak}
 V\zeta  &=& [V,\zeta]+ \zeta V;\nonumber \\
\vdots && \nonumber \\
V\zeta^k &=& \sum_{n=1}^k \zeta^{n-1}[V,\zeta] \zeta^{k-n}+\zeta^k V.
\end{eqnarray}
Also for any $n \geq 1$,
\begin{eqnarray}
\label{Vzetak1}
\zeta [V,\zeta] &=& [V,\zeta]\zeta + \mu; \nonumber \\
\vdots && \nonumber \\
 \zeta^{n-1} [V,\zeta]&=&[V,\zeta]\zeta^{n-1}
+ (n-1)\zeta^{n-2}\mu.
\end{eqnarray}
Therefore using (\ref{Vzetak}) and (\ref{Vzetak1}), it follows that
\begin{eqnarray*}
V\zeta^k &=&\sum_{n=1}^k [V,\zeta] \zeta^{n-1}\zeta^{k-n}+ (n-1)\zeta^{n-2}\zeta^{k-n}\mu  \nonumber \\
&&+\zeta^k V\nonumber \\
&=& \sum_{n=1}^k [V,\zeta] \zeta^{k-1}+ (n-1)\zeta^{k-2}\mu  +\zeta^k V\nonumber \\
&=&k[V,\zeta] \zeta^{k-1}+\frac{k(k-1)}{2}\zeta^{k-2}\mu+\zeta^k V
\end{eqnarray*}
which holds for any $k \geq 0$. 
Similarly
\begin{eqnarray*}
(\zeta^*)^k V &=& k(\zeta^*)^{k-1}[\zeta^*,V]+\frac{k(k-1)}{2}\mu^*(\zeta^*)^{k-2}
\nonumber \\
&&+V(\zeta^*)^k.
\end{eqnarray*}

Now given any $H_2 \in \mathcal{W}_4$, $k \geq 0$, $\ell \geq 0$,  we have
\begin{eqnarray}
\label{VHkl}
[V,H_{k\ell}] &=& k[V,\zeta] \zeta^{k-1}(\zeta^*)^\ell+\frac{k(k-1)}{2}\mu\zeta^{k-2}(\zeta^*)^\ell\nonumber \\&&
+\zeta^k V(\zeta^*)^\ell\nonumber \\
&& -k\zeta^\ell(\zeta^*)^{k-1}[\zeta^*,V]-\frac{k(k-1)}{2}\mu^*\zeta^\ell(\zeta^*)^{k-2}\nonumber \\&&
-\zeta^\ell V(\zeta^*)^k\nonumber \\
&=& k[V,\zeta] \zeta^{k-1}(\zeta^*)^\ell - k\zeta^\ell(\zeta^*)^{k-1}[\zeta^*,V]\nonumber \\
&&+\frac{k(k-1)}{2}\mu\zeta^{k-2}(\zeta^*)^\ell\nonumber \\
&&-\frac{k(k-1)}{2}\mu^*\zeta^\ell(\zeta^*)^{k-2}.
\end{eqnarray}
Therefore,
\begin{eqnarray}
\label{VH2}
[V,H_2] &=& \sum_{k=0}^\infty \sum_{\ell=0}^\infty S_{k\ell} [V,H_{k\ell}] \nonumber \\
&=& [V,\zeta]f'(\zeta,\zeta^*)-f'(\zeta,\zeta^*)^*[\zeta^*,V]\nonumber \\
&&+ \frac{1}{2}\mu f''(\zeta,\zeta^*)-\frac{1}{2}\mu^* f''(\zeta,\zeta^*)^*.
\end{eqnarray}
Now letting 
\begin{equation}
\label{zw1w2}
z = \zeta,~w_1 = f'(\zeta,\zeta^*)^*,\mbox{ and }w_2=  f''(\zeta,\zeta^*)^*,
\end{equation}
 it follows that condition (\ref{alt_comm_condition}) is
 satisfied. Furthermore, conditions (\ref{sector2a}), (\ref{sector2b})
 follow from conditions (\ref{sector4a}), (\ref{sector4b})
 respectively. Hence, $H_2 \in
\mathcal{W}_2$. Since, $H_2 \in
\mathcal{W}_4$ was arbitrary, we must have $\mathcal{W}_4 \subset \mathcal{W}_2$.
\hfill $\Box$

\section{The Linear  Case}
\label{sec:linear}
We now consider the  case in which the nominal quantum system corresponds to a linear quantum system; e.g., see \cite{JNP1,NJP1,MaP3,MaP4,PET10Ba}. In this case, we assume that $H_1$ is of the form 
\begin{equation}
\label{H1}
H_1 = \frac{1}{2}\left[\begin{array}{cc}a^\dagger &
      a^T\end{array}\right]M
\left[\begin{array}{c}a \\ a^\#\end{array}\right]
\end{equation}
where $M \in \mathbb{C}^{2n\times 2n}$ is a Hermitian matrix of the
form
\[
M= \left[\begin{array}{cc}M_1 & M_2\\
M_2^\# &     M_1^\#\end{array}\right]
\]
and $M_1 = M_1^\dagger$, $M_2 = M_2^T$.  In addition, we assume $L$ is of the form 
\begin{equation}
\label{L}
L = \left[\begin{array}{cc}N_1 & N_2 \end{array}\right]
\left[\begin{array}{c}a \\ a^\#\end{array}\right]
\end{equation}
where $N_1 \in \mathbb{C}^{m\times n}$ and $N_2 \in
\mathbb{C}^{m\times n}$. Also, we write
\[
\left[\begin{array}{c}L \\ L^\#\end{array}\right] = N
\left[\begin{array}{c}a \\ a^\#\end{array}\right] =
\left[\begin{array}{cc}N_1 & N_2\\
N_2^\# &     N_1^\#\end{array}\right]
\left[\begin{array}{c}a \\ a^\#\end{array}\right].
\]

In addition we assume that $V$ is of the form 
\begin{equation}
\label{quadV}
V = \left[\begin{array}{cc}a^\dagger &
      a^T\end{array}\right]P
\left[\begin{array}{c}a \\ a^\#\end{array}\right]
\end{equation}
where $P \in \mathbb{C}^{2n\times 2n}$ is a positive-definite Hermitian matrix of the
form
\begin{equation}
\label{Pform}
P= \left[\begin{array}{cc}P_1 & P_2\\
P_2^\# &     P_1^\#\end{array}\right].
\end{equation}
 Hence, we consider the set of  non-negative self-adjoint operators
$\mathcal{P}_1$ defined as
\begin{equation}
\label{P1}
\mathcal{P}_1 = \left\{\begin{array}{l}V \mbox{ of the form
      (\ref{quadV}) such that $P > 0$ is a 
} \\
\mbox{  Hermitian matrix of the form (\ref{Pform})}\end{array}\right\}.
\end{equation}

In the linear case, we will consider a specific notion of robust mean square stability. 
\begin{definition}
\label{D1}
An uncertain open quantum system defined by  $(S,L,H)$ where $H=H_1+H_2$ with $H_1$ of the form (\ref{H1}), $H_2 \in \mathcal{W}$, and $L$  of the form (\ref{L}) is said to be {\em robustly mean square stable} if for any $H_2 \in \mathcal{W}$, there exist constants $c_1 > 0$, $c_2 > 0$ and $c_3 \geq 0$ such that
\begin{eqnarray}
\label{ms_stable0}
\lefteqn{\left< \left[\begin{array}{c}a(t) \\ a^\#(t)\end{array}\right]^\dagger \left[\begin{array}{c}a(t) \\ a^\#(t)\end{array}\right] \right>}\nonumber \\
&\leq& c_1e^{-c_2t}\left< \left[\begin{array}{c}a \\ a^\#\end{array}\right]^\dagger \left[\begin{array}{c}a \\ a^\#\end{array}\right] \right>
+ c_3~~\forall t \geq 0.
\end{eqnarray}
Here $\left[\begin{array}{c}a(t) \\ a^\#(t)\end{array}\right]$ denotes the Heisenberg evolution of the vector of operators $\left[\begin{array}{c}a \\ a^\#\end{array}\right]$; e.g., see \cite{JG10}.
\end{definition}

In order to address the issue of robust mean square stability for the
uncertain linear quantum systems under consideration, we first require some algebraic identities.
\begin{lemma}
\label{L2}
Given $V \in \mathcal{P}_1$, $H_1$ defined as in (\ref{H1}) and $L$ defined as in (\ref{L}), then
\begin{eqnarray*}
\lefteqn{[V,H_1] =}\nonumber \\
&& \left[\left[\begin{array}{cc}a^\dagger &
      a^T\end{array}\right]P
\left[\begin{array}{c}a \\ a^\#\end{array}\right],\frac{1}{2}\left[\begin{array}{cc}a^\dagger &
      a^T\end{array}\right]M
\left[\begin{array}{c}a \\ a^\#\end{array}\right]\right] \nonumber \\
&=& \left[\begin{array}{c}a \\ a^\#\end{array}\right]^\dagger 
\left[
PJM - MJP 
\right] \left[\begin{array}{c}a \\ a^\#\end{array}\right].
\end{eqnarray*}


Also,
\begin{eqnarray*}
\lefteqn{\mathcal{L}(V) =} \nonumber \\
&& \frac{1}{2}L^\dagger[V,L]+\frac{1}{2}[L^\dagger,V]L \nonumber \\
&=& \tr\left(PJN^\dagger\left[\begin{array}{cc}I & 0 \\ 0 & 0 \end{array}\right]NJ\right)
\nonumber \\&&
-\frac{1}{2}\left[\begin{array}{c}a \\ a^\#\end{array}\right]^\dagger
\left(N^\dagger J N JP+PJN^\dagger J N\right)
\left[\begin{array}{c}a \\ a^\#\end{array}\right].
\end{eqnarray*}



Furthermore, 
\[
\left[\left[\begin{array}{c}a \\ a^\#\end{array}\right],\left[\begin{array}{cc}a^\dagger &
      a^T\end{array}\right]P
\left[\begin{array}{c}a \\ a^\#\end{array}\right]\right] = 2JP\left[\begin{array}{c}a \\ a^\#\end{array}\right].
\]
\end{lemma}
{\em Proof:}
The proof of these identities follows via  straightforward but tedious
calculations using (\ref{CCR2}). \hfill $\Box$

\subsection{Quadratic Hamiltonian Perturbations}
We now specialize the results of Section \ref{sec:quad_pert} to the case of a linear nominal system in order to obtain  concrete conditions for robust mean square stability. In this case, we use the relationship (\ref{z0}):
\begin{equation}
\label{z01}
z = \left[\begin{array}{c}\zeta \\ \zeta^\#\end{array}\right]= 
\left[\begin{array}{cc}
E_1 & E_2\\
E_2^\# & E_1^\#
\end{array}\right]\left[\begin{array}{c}a \\ a^\#\end{array}\right] 
=E\left[\begin{array}{c}a \\ a^\#\end{array}\right],
\end{equation}
to show that the following 
 following strict bounded real conditions 
provides a sufficient condition for robust mean square stability when $H_2 \in \mathcal{W}_3$: 
\begin{enumerate}
\item
The matrix 
\begin{equation}
\label{Hurwitz}
F = -iJM-\frac{1}{2}JN^\dagger J N\mbox{ is Hurwitz;}
\end{equation}
\item
\begin{equation}
\label{Hinfbound}
\left\|E\left(sI -F\right)^{-1}D \right\|_\infty < \frac{\gamma}{2}
\end{equation}
where $D = iJE^\dagger$.
\end{enumerate}

This leads to the following theorem.

\begin{theorem}
\label{T3}
Consider an uncertain open quantum system defined by $(S,L,H)$  such
that $H=H_1+H_2$ where $H_1$ is of the form (\ref{H1}), $L$ is of the
form (\ref{L}) and $H_2 \in \mathcal{W}_3$. Furthermore, assume that
the strict bounded real conditions  (\ref{Hurwitz}), (\ref{Hinfbound})
are satisfied. Then the uncertain quantum system is robustly mean square stable. 
\end{theorem}

\noindent
{\em Proof:}
If the conditions of the theorem are satisfied, then it follows from the strict bounded real lemma that the matrix inequality 
\begin{equation}
\label{QMI1}
F^\dagger P + P F +4PJE^\dagger EJP
+\frac{E^\dagger E}{\gamma^2}
 < 0.
\end{equation}
will have a solution $P > 0$ of the form (\ref{Pform}); e.g., see \cite{ZDG96,MaP4}.  This matrix $P$ defines a corresponding operator $V \in \mathcal{P}_1$ as in (\ref{quadV}). 
Now it follows from Lemma \ref{L2} that
\begin{eqnarray}
\label{zV}
[z,V] 
&=& 2E
JP\left[\begin{array}{c}a \\ a^\#\end{array}\right]
\end{eqnarray}
where $z$ is defined as in (\ref{z0}) and (\ref{z01}). Hence,
\[
[V,z^\dagger] [z,V] = 4\left[\begin{array}{c}a \\ a^\#\end{array}\right]^\dagger PJ 
E^\dagger E
JP
\left[\begin{array}{c}a \\ a^\#\end{array}\right].
\]
Also,
\[
z^\dagger z = \left[\begin{array}{c}a \\ a^\#\end{array}\right]^\dagger
E^\dagger E
\left[\begin{array}{c}a \\ a^\#\end{array}\right].
\]
Hence using Lemma \ref{L2}, we obtain
\begin{eqnarray}
\label{lyap_ineq2}
\lefteqn{-i[V,H_1]+ \mathcal{L}(V)+ [V,z^\dagger][z,V] + \frac{z^\dagger z}{\gamma^2} + cV} \nonumber \\
&=& \left[\begin{array}{c}a \\ a^\#\end{array}\right]^\dagger\left(\begin{array}{c}
F^\dagger P + P F\\ +4PJE^\dagger EJP
+\frac{E^\dagger E}{\gamma^2}
\end{array}\right)\left[\begin{array}{c}a \\ a^\#\end{array}\right]\nonumber \\
&&+\tr\left(PJN^\dagger\left[\begin{array}{cc}I & 0 \\ 0 & 0 \end{array}\right]NJ\right)\
\end{eqnarray}
where $F = -iJM-\frac{1}{2}JN^\dagger J N$. 
Therefore, it follows from (\ref{QMI1}) that there exists a constant $c > 0$ such that condition (\ref{dissip}) will be satisfied with 
\[
\tilde \lambda = \tr\left(PJN^\dagger\left[\begin{array}{cc}I & 0 \\ 0 & 0 \end{array}\right]NJ\right) \geq 0.
\]
 Hence choosing $\delta =0$, it follows from Lemma \ref{LA}, Theorem \ref{T1} and  $P > 0$ that
\begin{eqnarray*}
\lefteqn{\left< \left[\begin{array}{c}a(t) \\ a^\#(t)\end{array}\right]^\dagger \left[\begin{array}{c}a(t) \\ a^\#(t)\end{array}\right] \right>}\nonumber \\
 &\leq &
\frac{\left<V(t)\right>}{\lambda_{min}[P]} \nonumber \\
&\leq & e^{-ct}\frac{\left<V\right>}{\lambda_{min}[P]} + \frac{\tilde \lambda}{c\lambda_{min}[P]} \nonumber \\
&\leq&  e^{-ct}\left< \left[\begin{array}{c}a(0) \\ a^\#(0)\end{array}\right]^\dagger \left[\begin{array}{c}a(0) \\ a^\#(0)\end{array}\right] \right>\frac{\lambda_{max}[P]}{\lambda_{min}[P]}\nonumber \\
&&+ \frac{\tilde \lambda}{c\lambda_{min}[P]}~~\forall t \geq 0.
\end{eqnarray*}
 Hence, the condition (\ref{ms_stable0}) is satisfied with $c_1 = \frac{\lambda_{max}[P]}{\lambda_{min}[P]} > 0$, $c_2 = c > 0$ and $c_3 = \frac{\tilde \lambda}{c\lambda_{min}[P]} \geq 0$. 
\hfill $\Box$

\subsection{Non-quadratic Hamiltonian Perturbations}
We now specialize the results of Section \ref{sec:nonquadratic} to the case of a linear  nominal system. In this case, we define
\begin{eqnarray}
\label{z}
z &=& \zeta = E_1a+E_2 a^\# \nonumber \\
&=& \left[\begin{array}{cc} E_1 & E_2 \end{array}\right]
\left[\begin{array}{c}a \\ a^\#\end{array}\right] = \tilde E 
\left[\begin{array}{c}a \\ a^\#\end{array}\right]
\end{eqnarray}
where $z$ is assumed to be a scalar operator. Then, we  show that the following 
 following strict bounded real condition
provides a sufficient condition for robust mean square stability when $H_2 \in \mathcal{W}_4$: 
\begin{enumerate}
\item
The matrix 
\begin{equation}
\label{Hurwitz1}
F = -iJM-\frac{1}{2}JN^\dagger J N\mbox{ is Hurwitz;}
\end{equation}
\item
\begin{equation}
\label{Hinfbound1}
\left\|\tilde E^\# \Sigma\left(sI -F\right)^{-1}\tilde D \right\|_\infty < \frac{\gamma}{2}
\end{equation}
where $\tilde D = J\Sigma \tilde E^T$ and $\Sigma = \left[\begin{array}{cc} 0 & I\\
I &0 \end{array}\right].
$
\end{enumerate}

This leads to the following theorem.

\begin{theorem}
\label{T4}
Consider an uncertain open quantum system defined by $(S,L,H)$  such that
$H=H_1+H_2$ where $H_1$ is of the form (\ref{H1}), $L$ is of the
form (\ref{L}) and $H_2 \in \mathcal{W}_4$. Furthermore, assume that
the strict bounded real condition  (\ref{Hurwitz1}), (\ref{Hinfbound1})
is satisfied. Then the
uncertain quantum system is robustly mean square stable. 
\end{theorem}

In order to prove this theorem, we require the following lemma.
\begin{lemma}
\label{L4}
Given any $V \in \mathcal{P}_1$, then
\[
\mu = \left[z,[z,V]\right] = \left[z^*,[z^*,V]\right]^* = 
-\tilde E \Sigma JPJ\tilde E^T.
\]
which is a constant. Hence, the set of operators $\mathcal{P}_1$ satisfies Assumption \ref{A1}. 
\end{lemma}
{\em Proof:}
The proof of this result follows via a straightforward but tedious
calculation using (\ref{CCR2}). \hfill $\Box$

\noindent
{\em Proof of Theorem \ref{T4}.}
If the conditions of the theorem are satisfied, then it follows from the strict bounded real lemma that the matrix inequality 
\begin{equation}
\label{QMI2}
F^\dagger P + P F 
+4 PJ\Sigma \tilde E^T \tilde E^\# \Sigma JP 
+ \frac{1}{ \gamma^2}\Sigma \tilde E^T \tilde E^\# \Sigma
 < 0.
\end{equation}
will have a solution $P > 0$ of the form (\ref{Pform}); e.g., see \cite{ZDG96,MaP4}.  This matrix $P$ defines a corresponding operator $V \in \mathcal{P}_1$ as in (\ref{quadV}).

It follows from (\ref{z}) that we can write
\begin{eqnarray*}
z^* &=& E_1^\#a^\#+E_2^\# a=\left[\begin{array}{cc} E_2^\# & E_1^\# \end{array}\right]
\left[\begin{array}{c}a \\ a^\#\end{array}\right]\nonumber \\
&=&  \tilde E^\# \Sigma \left[\begin{array}{c}a \\ a^\#\end{array}\right].
\end{eqnarray*}
Also,  it follows from Lemma \ref{L2} that
\[
[z^*,V] = 2 \tilde E^\# \Sigma
JP\left[\begin{array}{c}a \\ a^\#\end{array}\right].
\]
Hence,
\begin{eqnarray}
\label{VzzV}
[V,z] [z^*,V] =
4\left[\begin{array}{c}a \\ a^\#\end{array}\right]^\dagger PJ 
\Sigma \tilde E^T \tilde E^\# \Sigma
JP
\left[\begin{array}{c}a \\ a^\#\end{array}\right].
\end{eqnarray}
Also, we can write
\begin{equation}
\label{zz}
zz^* = \left[\begin{array}{c}a \\ a^\#\end{array}\right]^\dagger
\Sigma \tilde E^T \tilde E^\# \Sigma
\left[\begin{array}{c}a \\ a^\#\end{array}\right].
\end{equation}

Hence using Lemma \ref{L2}, we obtain
\begin{eqnarray}
\label{lyap_ineq3}
&&-i[V,H_1]+ \mathcal{L}(V)
+ [V,z][z^*,V]
+\frac{zz^*}{\gamma^2}
 \nonumber \\
&=& \left[\begin{array}{c}a \\ a^\#\end{array}\right]^\dagger\left(\begin{array}{c}
F^\dagger P + P F\\ 
+4 PJ\Sigma \tilde E^T \tilde E^\# \Sigma JP \\
+ \frac{1}{ \gamma^2}\Sigma \tilde E^T \tilde E^\# \Sigma\\
\end{array}\right)\left[\begin{array}{c}a \\
a^\#\end{array}\right]\nonumber \\
&&+\tr\left(PJN^\dagger\left[\begin{array}{cc}I & 0 \\ 0 & 0 \end{array}\right]NJ\right)
\end{eqnarray}
where $F = -iJM-\frac{1}{2}JN^\dagger J N$. 

From this, it follows using (\ref{QMI2}) that there exists a constant $c > 0$ such that condition 
(\ref{dissip1a}) is satisfied with 
\[
\tilde \lambda = \tr\left(PJN^\dagger\left[\begin{array}{cc}I & 0 \\ 0 & 0 \end{array}\right]NJ\right) \geq 0.
\]
Hence, it follows from Lemma \ref{L4}, Lemma \ref{LB},  Theorem \ref{T2} and  $P > 0$ that 
\begin{eqnarray}
\label{ms_stable1}
\lefteqn{\left< \left[\begin{array}{c}a(t) \\ a^\#(t)\end{array}\right]^\dagger \left[\begin{array}{c}a(t) \\ a^\#(t)\end{array}\right] \right>}\nonumber \\
&\leq&  e^{-ct}\left< \left[\begin{array}{c}a(0) \\ a^\#(0)\end{array}\right]^\dagger \left[\begin{array}{c}a(0) \\ a^\#(0)\end{array}\right] \right>\frac{\lambda_{max}[P]}{\lambda_{min}[P]}\nonumber \\
&&+ \frac{\lambda}{c\lambda_{min}[P]}~~\forall t \geq 0
\end{eqnarray} 
where $\lambda = \tilde \lambda+ \delta_1+\mu\mu^*/4+\delta_2.$
 Hence, the condition (\ref{ms_stable0}) is satisfied with $c_1 = \frac{\lambda_{max}[P]}{\lambda_{min}[P]} > 0$, $c_2 = c > 0$ and $c_3 = \frac{\lambda}{c\lambda_{min}[P]} \geq 0$. 
\hfill $\Box$

\section{Illustrative Example}
\label{sec:example}
We  consider an example of an open quantum system with 
\[
S=I,~H_1=0,~H_2 = \frac{1}{2}i\left(\left(a^\dagger\right)^2-a^2\right),~L
= \sqrt{\kappa}a,
\]
which corresponds  an optical
parametric amplifier; see \cite{GZ00}. This defines a linear quantum
system of the form considered in Theorem \ref{T3} with $M_1 = 0$, $M_2 =
0$, $N_1 = \sqrt{\kappa}$, $N_2 = 0$, $E_1=1$, $E_2 = 0$, $\Delta_1 =
0$, $\Delta_2 = i$. 
Hence, $M=0$, $N=\left[\begin{array}{ll}\sqrt{\kappa} & 0\\0 &
    \sqrt{\kappa}\end{array}\right]$, $F= \left[\begin{array}{ll}-\frac{\kappa}{2} & 0\\0 &
    -\frac{\kappa}{2}\end{array}\right]$ which is Hurwitz, $E=I$, and
$D = iJ$.
In this case, 
\[
\Delta \Delta = \left[\begin{array}{cc}0 & i\\
-i &     0\end{array}\right]\left[\begin{array}{cc}0 & i\\
-i &     0\end{array}\right] =\left[\begin{array}{cc}1 & 0\\
0 &     1\end{array}\right].
\]
Hence, we can choose $\gamma = 1$ to ensure that (\ref{Delta_bound})
is satisfied and $H_2 \in \mathcal{W}_3$. Also, 
\[
\left\|E\left(sI -F\right)^{-1}D \right\|_\infty 
= \left\|\left[\begin{array}{ll}\frac{1}{s+\kappa/2} & 0\\0 &
    -\frac{1}{s+\kappa/2}\end{array}\right]\right\|_\infty =\frac{2}{\kappa}.
\]
Hence, it follows from Theorem \ref{T3} that this system will be mean
square stable if $\frac{2}{\kappa} < \frac{1}{2}$; i.e., $\kappa >
4$. 

\section{Conclusions}
\label{sec:conclusions}
In this paper, we have considered the problem of robust stability for
uncertain  quantum systems with either quadratic and non-quadratic
perturbations to the system Hamiltonian. The final stability results
obtained are expressed in terms of strict bounded real 
conditions. Future research will be directed towards analyzing the stability of specific nonlinear quantum systems
using the given robust stability results for the case of non-quadratic
perturbations to the system Hamiltonian.


\end{document}